\documentclass{sig-alternate-05-2015}

\makeatletter
\newif\if@restonecol
\makeatother

\usepackage[T1]{fontenc}

\usepackage{amsmath}
\usepackage{etex}
\usepackage[usenames,dvipsnames,table]{xcolor}
\usepackage[pdfpagelabels=false,draft]{hyperref}
\usepackage{listings}
\usepackage{booktabs}
\usepackage{multirow}
\usepackage{fixltx2e} 
\usepackage[font=bf]{caption}
\usepackage{mathpartir} 
\usepackage[ruled, vlined, linesnumbered]{algorithm2e}
\usepackage{mathpazo}
\usepackage{subcaption}
\usepackage{pifont} 
\usepackage{pgf-pie}
\usepackage{pgfplots}
\usepgfplotslibrary{groupplots}
\usepackage{pgfplotstable}
\usepackage{tikz}
\usepackage{afterpage}
\usepackage{xcolor}
\hypersetup{
    colorlinks,
    linkcolor={red!50!black},
    citecolor={blue!50!black},
    urlcolor={blue!80!black}
}
\usetikzlibrary{decorations.markings,decorations.pathmorphing}
\usetikzlibrary{patterns, matrix, positioning, fit, backgrounds, arrows, shapes}
\usepackage{times}

\usepackage{amsthm}
\definecolor{skyblue1}{rgb}{0.447,0.624,0.812}
\pgfplotsset{width=12cm,compat=newest}

\graphicspath{{experiment/}}

\newcommand{\code}[1]{\texttt{\small {#1}}}
\newcommand{\summary}[1]{\texttt{#1}}

\newcommand{\nosemicolon}{\renewcommand{\@endalgocfline}{\relax}}
\newcommand{\dosemicolon}{\renewcommand{\@endalgocfline}{\algocf@endline}}

\definecolor{lgray}{gray}{0.8}

\tikzstyle{decision} = [diamond, draw, fill=blue!20, 
    text width=4.5em, text badly centered, node distance=3cm, inner sep=0pt]
\tikzstyle{refimpl} = [rectangle, draw, fill=blue!20, 
    text width=4em, text centered, rounded corners, minimum height=3em]
\tikzstyle{sub} = [rectangle, draw=lightgray, fill=red!20, 
    text width=4em, text centered, rounded corners, minimum height=3em]
\tikzstyle{sum} = [rectangle, draw, fill=white,
    text width=4em, text centered, rounded corners, minimum height=3em]
\tikzstyle{block} = [rectangle, draw, fill=black!20, 
    text width=4em, text centered, minimum height=4em]
\tikzstyle{cloud} = [draw, ellipse,fill=green!20, node distance=3cm,
    text centered, text width=1.4cm, minimum height=2em]
\tikzstyle{widecloud} = [draw, ellipse,fill=green!20, node distance=3cm,
    text centered, text width=1.8cm, minimum height=2em]
\tikzstyle{vecArrow} = [thick, decoration={markings,mark=at position
   1 with {\arrow[semithick]{open triangle 60}}},
   double distance=1.4pt, shorten >= 5.5pt,
   preaction = {decorate},
   postaction = {draw,line width=1.4pt, white,shorten >= 4.5pt}]

\pgfplotsset{
   /pgfplots/bar  cycle  list/.style={/pgfplots/cycle  list={%
        {blue,fill=blue!50,mark=none},%
        {red,fill=red!50,mark=none},%
        {brown,fill=brown!50,mark=none},%
        {green,fill=green!50,mark=none},%
     }
   },
}

\newcounter{theexample}

\SetKwData{iter}{iter}
\SetKwData{body}{body}
\SetKwData{guard}{guard}
\SetKwData{updated}{updated}
\SetKwData{used}{used}
\SetKwData{idx}{idx}
\SetKwData{iVar}{iVar}
\SetKwData{pre}{Pre}
\SetKwData{post}{Post}
\SetKwData{missing}{Missing}
\SetKwData{spurious}{Spurious}
\SetKwData{equi}{Equiv}

\SetKwData{refBlock}{$\mathcal{B}_R$}
\SetKwData{subBlock}{$\mathcal{B}_S$}
\SetKwData{varMap}{$\rho$}
\SetKwData{inputConstraints}{IC}

\newtheoremstyle{mystyle}
  {}
  {}
  {}
  {}
  {\bfseries}
  {.}
  { }
  {\thmname{#1}\thmnumber{ #2}\thmnote{ (#3)}}

\theoremstyle{mystyle}

\theoremstyle{mystyle}

\theoremstyle{mystyle}

\def\denseitems{
  \itemsep1pt plus1pt minus1pt
  \parsep0pt plus0pt
  \parskip0pt\topsep0pt}

\let\oldnl\nl
\newcommand{\nonl}{\renewcommand{\nl}{\let\nl\oldnl}}
\makeatother

\SetCommentSty{mycommfont}

\newcommand{\ignore}[1]{}

\begin{document}

  
\setcopyright{acmcopyright}
 



\makeatletter
\def\@copyrightspace{\relax}
\makeatother

\title{Semi-Supervised Verified Feedback Generation}

\numberofauthors{1}
\author{
\begin{tabular}{cc}
\begin{tabular}{c}
\begin{tabular}{ccc}
  Shalini Kaleeswaran & Anirudh Santhiar & Aditya Kanade\\
\end{tabular}\\
\affaddr{Indian Institute of Science, Bangalore}\\
\email{\{shalinik,anirudh\_s,kanade\}@csa.iisc.ernet.in}\\
\end{tabular}
&
\begin{tabular}{c}
  Sumit Gulwani\\
  \affaddr{Microsoft Research, Redmond}\\
  \email{sumitg@microsoft.com}
\end{tabular}
\end{tabular}
}

\maketitle

\begin{abstract}
  Students have enthusiastically
  taken to online programming lessons and contests.
  Unfortunately, they tend to struggle due to lack of personalized feedback when
  they make mistakes. The overwhelming number of submissions
  precludes manual evaluation. There is an urgent need of 
  program analysis and repair techniques capable of
  handling both the scale and variations in student submissions, while
  ensuring quality of feedback.

  Towards this goal, we present a novel methodology called \emph{semi-supervised
  verified feedback generation}. We cluster submissions by solution
  strategy and ask the instructor to identify or add a correct submission in each
  cluster. We then verify every submission in a cluster against the
  instructor-validated submission in the same cluster. If faults are
  detected in the submission then feedback suggesting fixes to them is generated.
  Clustering reduces
  the burden on the instructor and also the variations that have to be handled
  during feedback generation. The verified feedback generation ensures that only
  correct feedback is generated.

  We have applied this methodology to iterative dynamic programming (DP)
  assignments. Our clustering technique uses features of DP solutions.
  We have designed a novel \emph{counter-example guided feedback generation}
  algorithm capable of suggesting fixes to all faults in a submission.
  In an evaluation on $2226$ submissions to $4$
  problems, we could generate \emph{verified} feedback for $1911$ ($85\%$)
  submissions in $1.6$s each on an average. Our technique does a good job of reducing the burden on the
  instructor. Only one submission had
  to be manually validated or added for every $16$ submissions. 
  \end{abstract}

\section{Introduction}
\label{sec:introduction}

Programming has become a much sought-after skill for superior employment in today's
technology-driven world~\cite{acm-edu}.
  Students have enthusiastically
  taken to online programming lessons
  and contests, in the hope of learning and improving programming skills.
  Unfortunately, they tend to struggle due to lack of personalized feedback when
  they make mistakes. The overwhelming number of student submissions
  precludes manual evaluation. There is an urgent need of automated
  program analysis and repair techniques capable of
  handling both the scale and variations in student submissions, while
  ensuring quality of feedback.

  A promising direction
is to cluster submissions, so that the instructor
provides feedback for a representative from each cluster which is 
then propagated automatically to other submissions in the same
cluster~\cite{HuangAIED,DBLP:conf/icml/PiechHNPSG15}.
This provides scalability while keeping the instructor efforts manageable.
Many novel solutions have been proposed in recent times to enable
clustering of programs.
These include syntactic or test-based similarity~\cite{gross2012cluster,Rivers,HuangAIED,Glassman:2014:FEC:2556325.2567865,overcode},
co-occurrence of code phrases~\cite{codewebs} and vector representations
obtained by deep learning~\cite{peng2015building,DBLP:conf/icml/PiechHNPSG15,mou2016convolutional}.
However, clustering can be correct only in a \emph{probabilistic} sense.
Thus, these techniques cannot 
guarantee that the feedback provided manually by the instructor,
by looking only at some submissions in a cluster, would indeed be suitable 
to all the submissions in that cluster. As a result, some submissions may receive incorrect
feedback. Further, if submissions that have similar mistakes
end up in different clusters, some of them may not receive the suitable feedback.
Instead of helping, these drawbacks can cause confusion among the students.

\begin{figure}
  \centering
  \setlength{\abovecaptionskip}{5pt}
  \setlength{\belowcaptionskip}{-15pt}
\begin{tikzpicture}[thick,scale=0.6]

\draw(2,1) ellipse (2 and 1);
\node[above, font=\scriptsize] at (2,2) {Student submissions};

\newcommand{\coords}{.5/1/1/1.5/2/1.8/2/1.2/2/.6/3/1.5/3,.5/3.5/1}

\newcounter{i}
\setcounter{i}{0}
\foreach \point in {(.5,1),(1,1.5),(1,.5),(2,1.6),(2,1),(2,.4),(3,1.5),(3,.5),(3.5,1)}
{
    \node (point-\arabic{i}) at \point {?};
    \stepcounter{i}
}

\draw(9,.5) ellipse (1 and .5);
\draw(11.5,.5) ellipse (1 and .5);
\draw(10.25,1.5) ellipse (1 and .5);
\node[above, font=\scriptsize] at (10.25,2) {Clusters of submissions};

\foreach \point in {(8.45,.4),(9.55,.4),(9,.65), (11,.4),(12,.4),(11.45,.65), (9.65,1.35), (10.25,1.65) ,(10.85,1.35)}
{
    \node (point-\arabic{i}) at \point {?};
    \stepcounter{i}
}

\draw(9,-2.5) ellipse (1 and .5);
\draw(11.5,-2.5) ellipse (1 and .5);
\draw(10.25,-3.5) ellipse (1 and .5);
\node[below, font=\scriptsize] at (10.25,-4) {Clusters with validated submissions};

\foreach \point in {(8.45,-2.4),(9,-2.65), (11,-2.4),(12,-2.4),(9.65,-3.35),(10.25,-3.65)}
{
    \node (point-\arabic{i}) at \point {?};
    \stepcounter{i}
}

\foreach \point in {(9.55,-2.4),(11.45,-2.65),(10.85,-3.35) }
{
    \node (point-\arabic{i}) at \point {$\color{green}\checkmark$};
    \stepcounter{i}
}


\draw(1,-2.5) ellipse (1 and .5);
\draw(3.5,-2.5) ellipse (1 and .5);
\draw(2.25,-3.5) ellipse (1 and .5);
\node[below, font=\scriptsize] at (2.25,-4) {Submissions with feedback};

\newcommand{\Cross}{\color{red}\ding{55}}%

\foreach \point in {(1,-2.65),(4,-2.4),(3.45,-2.65), (2.85,-3.35),(1.55,-2.4),(3,-2.4)}
{
    \node (point-\arabic{i}) at \point {$\color{green}\checkmark$};
    \stepcounter{i}
}

\foreach \point in {(0.45,-2.5),(1.65,-3.55) }
{
    \node (point-\arabic{i}) at \point {\Cross};
    \stepcounter{i}
}

\foreach \point in {(2.25,-3.65)}
{
    \node (point-\arabic{i}) at \point {?};
    \stepcounter{i}
}

\draw[-latex] (4.5,1) -- node[above, font=\scriptsize] {Clustering by} node[below, font=\scriptsize] {solution strategy} (8,1); 
\draw[-latex] (10.25,0) -- (10.25,-2) node[midway, right, align=center, font=\scriptsize] {Instructor-\\validated\\submissions}; 
\draw[-latex] (8,-3) -- node[above, font=\scriptsize] {Verified feedback} node[below, font=\scriptsize] {generation} (4.5,-3);  
\end{tikzpicture}
\caption{Semi-supervised verified feedback generation:
{\color{green}\ding{51}} is a submission verified to be correct,
{\color{red}\ding{55}} is a faulty submission for which feedback is generated
and ? is an unlabeled submission for which feedback is not generated.}
\label{fig:overview}
\end{figure}
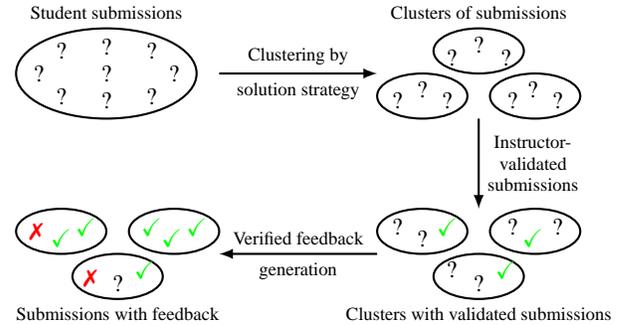

To overcome these drawbacks, we propose a novel methodology in which
clustering of submissions is followed by
an automated feedback generation phase grounded in
formal verification. Figure~\ref{fig:overview} shows our 
methodology, called \emph{semi-supervised verified
feedback generation}. Given a set of unlabeled
student submissions, we first cluster them by similarity of
solution strategies and ask
the instructor to identify\footnote{To minimize
  instructor's efforts further, we discuss
  strategies to suggest potentially correct submissions to the instructor.}
  a correct submission in
each cluster. If none exists, the instructor adds a correct
solution similar to the submissions in the cluster;
after which we do clustering again. 
In the next phase, each submission in a cluster is
\emph{verified against the instructor-validated submission} in the same cluster. 
If any faults are detected in
the submission then feedback suggesting fixes to them is generated.
Because program equivalence checking is an undecidable problem,
it may not be possible to generate feedback for every submission.
We let the instructor evaluate such submissions manually.
This is better than 
propagating unverified or incorrect feedback indiscriminately.

The proposed methodology has several advantages. First, it uses unsupervised
clustering to reduce the burden on the instructor. The supervision from the
instructor comes in the form of identifying 
a correct submission per cluster. 
Second, the verification phase complements clustering by
providing \emph{certainty} about correctness of feedback.
Clustering helps by reducing the variations required to be handled during
feedback generation.
It would be difficult to check equivalence of very dissimilar submissions,
\emph{e.g.}, an iterative solution against a recursive solution.
Since only submissions that are similar
(\emph{i.e.}, belong to the same cluster) are compared,
there is a higher chance of making equivalence checking work in practice.
Third, feedback is generated for each submission
separately so it is personalized.
This is superior to propagating a
manually-written feedback indiscriminately to all the submissions in a cluster.

\begin{figure}[t]
\setlength{\belowcaptionskip}{-15pt}
\small{
\begin{lstlisting}
void main() {
  int i, j, n, max; 
  (*@\textsf{scanf}@*)("%d", &n);    // Input (*@\label{line:inp1}@*)
  int m[n][n], dp[n][n]; // dp is the DP array (*@\label{line:dpdecl}@*)
  for (i = 0; i < n; i++) (*@\label{line:inp2loopstart}@*)
    for (j = 0; j <= i; j++)
      (*@\textsf{scanf}@*)("%d", &m[i][j]); // Input (*@\label{line:inp2}@*)
  dp[0][0] = m[0][0]; // Initialization (*@\label{line:init}@*)
  for (i = 1; i < n; i++) { (*@\label{line:outerloopstart}@*)
    for (j = 0; j <= i; j++) { (*@\label{line:inlpstart}@*)
      if (j == 0)   (*@\label{line:updt1guard}@*) 
        dp[i][j] = dp[i-1][j] + m[i][j]; // Update (*@\label{line:updt1}@*)
      else if (j == i) (*@\label{line:updt2guard}@*)
        dp[i][j] = dp[i-1][j-1] + m[i][j]; // Update (*@\label{line:updt2}@*)
      else if (dp[i-1][j] > dp[i-1][j-1]) (*@\label{line:updt3guard}@*)
        dp[i][j] = dp[i-1][j] + m[i][j]; // Update (*@\label{line:updt3}@*)
      else dp[i][j] = dp[i-1][j-1] + m[i][j]; // Update (*@\label{line:updt4}@*)
    } (*@\label{line:inlpend}@*)
  } (*@\label{line:outerloopend}@*)
  max = dp[n-1][0]; (*@\label{line:maxcomputestart}@*)
  for (i = 1; i < n; i++)
    if (dp[n-1][i] > max) max = dp[n-1][i]; (*@\label{line:maxcomputeend}@*)
  (*@\textsf{printf}@*)("%d", max); // Output (*@\label{line:output}@*)
}
\end{lstlisting}
}
\caption{A correct submission for the matrix path problem.}
\label{fig:matrixpath}
\end{figure}

\begin{figure}[t]
\small{
\begin{lstlisting}
int max(int a, int b) {
  return a > b ? a : b;
}
int max_arr(int arr[]) {
  int i, max;
  max = arr[0]; (*@\label{subline:max1}@*)
  for (i = 0; i < 100; i++)
    if (arr[i] > max) max = arr[i]; (*@\label{subline:max2}@*)
  return max;
}
int main() {
  int n, i, j, A[101][101], D[101][101]; // D is the DP array
  (*@\textsf{scanf}@*)("%d", &n);// Input (*@\label{subline:inp1}@*)
  for (i = 0; i < n; i++) (*@\label{subline:inploopstart}@*)
    for (j = 0; j <= i; j++)
      (*@\textsf{scanf}@*)("%d", &A[i][j]); // Input (*@\label{subline:inp2}@*)
  D[0][0] = A[0][0] (*@\label{subline:init}@*); // Initialization
  for (i = 1; i < n; i++) (*@\label{subline:updtloopstart}@*)
    for (j = 0; j <= i; j++) (*@\label{subline:innloopstart}@*)
      D[i][j] = A[i][j] + max(D[i-1][j], D[i-1][j-1]); // Update (*@\label{subline:update}@*)
  int ans = max_arr(D[n-1]); (*@\label{subline:output}@*) 
  (*@\textsf{printf}@*)("%d", ans); // Output (*@\label{subline:output}@*)
  return 0;
}
\end{lstlisting}
}
\caption{A faulty submission belonging to the same cluster as
  the correct submission in Figure~\ref{fig:matrixpath}.}
\label{fig:faultymatrix}
\end{figure}

To demonstrate the effectiveness of this methodology,
we apply it to iterative dynamic programming assignments.
Dynamic programming (DP)~\cite{Bellman:2003:DP:862270} is a standard 
technique taught in algorithms courses. Shortest-path and subset-sum
are among the many problems that can be solved efficiently by DP.
We design features that characterize the DP strategy
and extract them from student submissions by static analysis
and pattern matching. The features include the types of the DP arrays
used for memoizing solutions to sub-problems, and how the sub-problems
are solved and reused iteratively. The set of features we need is small
and all features are discrete valued.
Therefore, our clustering approach is very simple
and works directly by checking equality of feature values.

We also propose a novel feedback generation algorithm called
\emph{counter-example guided feedback generation}. The
equivalence between two submissions is checked using syntactic simplifications
and satisfiability-modulo-theories (SMT) based constraint solving.
If an equivalence check fails, our algorithm uses the counter-example
generated by the SMT solver to refine the equivalence query.
This process terminates when our algorithm
proves the equivalence, or is unable to refine the query.
A trace of refinements leading to a logically valid
equivalence query constitutes the feedback.

As an example, consider the matrix path
problem\footnote{\small{\url{http://www.codechef.com/problems/SUMTRIAN}}} taken
from a popular programming contest site \href{http://codechef.com}{CodeChef}.
A lower triangular matrix of \code{n} rows is given.  Starting at a
cell, a path can be traversed in the matrix by moving either directly below or
diagonally below to the right. The objective is to find the maximum weight among
the paths that start at the cell in the first row and first column, and end in
any cell in the last row. The weight of a path is the sum of all cells along
that path. Figure~\ref{fig:matrixpath} shows 
an example correct submission to this problem
and Figure~\ref{fig:faultymatrix} shows a faulty submission. The two programs
are syntactically and structurally quite different but
both use 2D integer arrays for memoization and
iterate over them in the same manner. Our clustering
technique therefore puts them into the same cluster.
This avoids unnecessarily creating many clusters but requires a more
powerful feedback generation algorithm that can handle
stylistic variations between submissions, \emph{e.g.}, 
Figure~\ref{fig:faultymatrix} uses multiple procedures whereas
Figure~\ref{fig:matrixpath} uses only one.

\begin{figure}[t]
\setlength{\abovecaptionskip}{1pt}
\setlength{\belowcaptionskip}{-15pt}
\small{
\begin{tabular}{l}
\textbf{In the declaration:} 1) Types of \code{A} and \code{D} should be \code{int[n][n]}\\
\textbf{In the update:}\\ 
2) Under guard \code{j == 0}, \\
\phantom{xxx}compute \code{D[i][j] = D[i-1][j] + A[i][j]}\\
\phantom{xxx}instead of \code{D[i][j] = A[i][j] + (D[i-1][j]>D[i-1][j-1] ?}\\
\phantom{xxxxxxxxxxxxxxxxxxxxx}\code{D[i-1][j] : D[i-1][j-1])}.\\
3) Under guard \code{(j != 0 \&\& j == i)}, \\
\phantom{xxx}compute \code{D[i][j] = D[i-1][j-1] + A[i][j]}\\
\phantom{xxx}instead of \code{D[i][j] = A[i][j] + (D[i-1][j]>D[i-1][j-1] ?}\\
\phantom{xxxxxxxxxxxxxxxxxxxxx}\code{D[i-1][j] : D[i-1][j-1])}.\\
\textbf{In the output:}
4) Under guard \code{true}, compute maximum\\
\phantom{xxx}
over \code{D[n-1][0],...,D[n-1][n-1]} instead of \code{D[n-1][0],...,D[n-1][99]}.
\end{tabular}
}
\caption{The auto-generated feedback for the submission
  in Figure~\ref{fig:faultymatrix} by verifying it against the
  submission in Figure~\ref{fig:matrixpath}.}
\label{fig:feedback}
\end{figure}

Our algorithm automatically generates the feedback in Figure~\ref{fig:feedback}
for the faulty submission by verifying it against the correct submission.
The first correction suggests that the submission should use array sizes as
\code{int[n][n]} instead of hardcoded value of \code{int[101][101]}.  The 
update to the DP array \code{D}
at line~\ref{subline:update} in Figure~\ref{fig:faultymatrix} misses some
corner cases for which our algorithm generates corrections \#2 and \#3 above.  The
computation of output at line~22 should use the correct array bounds as
indicated by correction \#4.
This is a comprehensive list of changes to correct the faulty submission.

We have implemented our technique for C programs and evaluated it on $2226$
student submissions to $4$ problems from CodeChef. 
On $1911$ ($85$\%) of them, we could generate feedback
by either verifying them to be correct, or identifying faults and fixes for them. 
In addition to faults in wrong answers,
we also found faults
in $265$ submissions accepted by CodeChef as {correct answers}!
Like most online contest sites, CodeChef uses test-based evaluation.
Our static verification technique has a qualitative advantage over the test-based approach 
of online judges. The submissions come from $1860$ 
students from over $250$ different institutes and are therefore
representative of diverse backgrounds and coding styles. Even then,
the number of clusters ranged only from $2$--$80$ across the problems.
Our technique does a good job of reducing the burden on the instructor.
On an average, using one manually validated or added
submission, we generated verified feedback on $16$
other submissions. We had to add only $7$ correct solutions manually.
While our technique generated feedback automatically for $1911$
submissions, the remaining $315$ ($15$\%) submissions require manual evaluation.
Our technique is fast and
on an average, took $1.6$s to generate feedback for each submission.

Work on feedback generation so far has focused on {\em introductory programming}
assignments~\cite{autograder,perfFeed,SrikantKDD,Ihantola}. 
In comparison, we address the challenging class of \emph{algorithmic} assignments,
in particular, that of dynamic programming.
The program repair approaches for developers~\cite{legoues-tse2012,nguyen2013,minthint,directfix,staged-program-repair}
deal with one program at a time.
We work with all student submissions simultaneously. To do so,
we propose a methodology inspired by both machine learning and verification.
Unlike the developer setting, we have the luxury of
calling upon the instructor to identify or add correct solutions. We exploit
this to give complete and correct feedback but then our technique must solve the 
challenging (and in general, undecidable)
problem of checking semantic equivalence of programs.

The salient contributions of this work are as follows:\\[-5mm]
\begin{itemize}\denseitems
\item We present a novel methodology of clustering of submissions followed
  by program equivalence checking within each cluster.
  This methodology can pave way for practical feedback generation
  tools capable of handling both the scale
  and variations in student submissions, while minimizing the instructor's
  efforts and ensuring quality of feedback. 
\item We demonstrate that this methodology is effective by
  applying it to the challenging class of iterative
  DP solutions. We design a clustering technique and
  a counter-example guided feedback generation algorithm for DP solutions.
\item We experimentally evaluate the technique on $2226$ submissions
  to $4$ problems and generate \emph{verified} feedback for $85\%$
  of them. We show that our technique does not require many inputs from
  the instructor and runs efficiently.
  \end{itemize}

\section{Detailed Example}
\label{sec:overview}

We now explain in details how our technique handles
the motivating example from the previous section.

\subsection{Clustering Phase}
\label{sec:clustering-example}

The two submissions in Figure~\ref{fig:matrixpath} and
Figure~\ref{fig:faultymatrix} are syntactically and structurally
quite different. 
Our technique extracts features of the solution
strategy in a submission. These features are more abstract than
low-level syntactic or structural features and put the superficially dissimilar
submissions into the same cluster.

The solution strategy of a DP program is characterized by the DP recurrence
being solved~\cite{Bellman:2003:DP:862270}. The DP recurrence for the correct submission
in Figure~\ref{fig:matrixpath} is as follows:
\[
\texttt{dp[i][j]} = 
\begin{cases}
  \texttt{m[0][0]} & \text{if}\ \texttt{i}\!=\!\texttt{0}, \texttt{j}\!=\!\texttt{0}\\
  \texttt{dp[i-1][j]+m[i][j]} & \text{if}\ \texttt{i} \!\neq\! \texttt{0},
  \texttt{j}\!=\!\texttt{0}\\
  \texttt{dp[i-1][j-1]+m[i][j]} & \text{if}\ \texttt{j}\!=\!\texttt{i}\!\neq\!\texttt{0}\\
  \texttt{max(dp[i-1][j],dp[i-1][j-1])+m[i][j]} & \text{otherwise}
\end{cases}
\]
where \code{dp} is the DP array, \code{m} is the input matrix
of \code{n} rows, and \code{i} goes from $0$ to \code{n-1},
\code{j} goes from $0$ to \code{i} and \code{max} returns the maximum of the
two numbers. The DP recurrence of the submission
in Figure~\ref{fig:faultymatrix} is similar but misses the second
and third cases above.

Comparing the recurrences directly would be ideal but
extracting them is not easy.
Students can implement a recurrence formula in different imperative styles.
They may use multiple procedures
(as in Figure~\ref{fig:faultymatrix}) and arbitrary
temporary variables to hold intermediate results.
Rather than attempting to extract the precise recurrence, we
extract some features of the solution to
find submissions that use similar solution strategies. For this, our
analysis identifies and labels the DP arrays used in each submission.
It also identifies and labels {key statements} that
1) read inputs, 2) initialize the DP array,
3) update the DP array elements using previously computed array elements,
and 4) generate the output. 
The comments in Figure~\ref{fig:matrixpath} and Figure~\ref{fig:faultymatrix}
identify the DP arrays and key statements.

We call a loop which is not contained within any other statement as a
\emph{top-level loop}. For example, the loop at
lines~\ref{line:outerloopstart}--\ref{line:outerloopend} in Figure~\ref{fig:matrixpath}
is a top-level loop but the loop at lines~\ref{line:inlpstart}--\ref{line:inlpend}
is not. More generally, a statement which is not contained
within any other statement is a \emph{top-level statement}.
The \emph{features extracted by our technique}
and their values for the submission in Figure~\ref{fig:matrixpath}
are as follows:\\[-5mm]
\begin{enumerate}\denseitems
\item Type and the number of dimensions of the DP array: $\langle \code{int}, 2 \rangle$
\item Whether the input array is reused as the DP array: No
\item The number of top-level loops which contain update statements
  for the DP array: $1$
\item For each top-level loop containing updates to the DP array,
  \begin{enumerate}
  \item The loop nesting depth: $2$
  \item The direction of loop indices: $\langle +,+ \rangle$ (indicating that
    the respective indices are incremented by one in each iteration of the
    corresponding loops)
  \item The DP array element updated inside the loop:
    \texttt{dp[i][j]}
  \end{enumerate}
\end{enumerate}


\newcommand{\dOneText}{{Using the counter-example $E_1$, the algorithm discovers that when
    $\texttt{j}=\texttt{0}$, $\varphi_2$ computes statement $s_1$ (line~\ref{subline:update}
    in Figure~\ref{fig:faultymatrix}) but according to
    $\varphi_1$, it should compute $s_2$:
    \texttt{D[i][j] = D[i-1][j]+A[i][j]}.\\[2mm]    
    Let $\varphi_2' \equiv (\texttt{j} \neq \texttt{0} \implies s_1)
    \wedge (\texttt{j} = \texttt{0} \implies s_2)$.    
}}

\newcommand{\dTwoText}{{Using the counter-example $E_2$, the algorithm discovers that when
    $\texttt{j}=\texttt{i} \wedge \texttt{j} \neq \texttt{0}$,
    $\varphi_2'$ computes $s_1$ but according to
    $\varphi_1$, it should compute $s_3$:
        \texttt{D[i][j] = D[i-1][j-1]+A[i][j]}.\\[2mm]    
   Let $\varphi_2'' \equiv
   (\texttt{j} \neq \texttt{0} \wedge \texttt{j} \neq \texttt{i} \implies s_1) \wedge
   (\texttt{j} \neq \texttt{0} \wedge \texttt{j} = \texttt{i} \implies s_3) \wedge
    (\texttt{j} = \texttt{0} \implies s_2)$.
}}

\begin{figure*}[t]
  \centering
  \setlength{\abovecaptionskip}{-28pt}
  \setlength{\belowcaptionskip}{-15pt}
  \scalebox{0.9}{
\begin{tikzpicture}[-latex]
  \node[draw, fill = gray!30, diamond, text centered, text width=1cm] (n1) {Is $\psi_1$ valid?};
  
  \node[draw, fill = gray!30, diamond, below left = 1.4cm and -.4 cm of n1, text centered, text width=1cm] (n2) {Is $\psi_2$ valid?};
  \node[color=white, diamond, below right = 1.4cm and -.4 cm of n1] (n4) {\phantom{Is $\psi_2$ valid?}};
  
  \node[draw, fill = gray!30, diamond, below left = 1.4cm and -.4 cm of n2, text centered, text width=1cm] (n3) {Is $\psi_3$ valid?};
  \node[color=white, diamond, below right = 1.4cm and -.4 cm of n2] (n5) {\phantom{Is $\psi_2$ valid?}};

  \node[color=white, diamond, below left = 1.4cm and -.4 cm of n3] (n6) {\phantom{Is $\psi_2$ valid?}};
  \node[color=white, diamond, below right = 1.4cm and -.4 cm of n3] (n7) {\phantom{Is $\psi_2$ valid?}};

  \node[below=-1.8cm of n7.south, text width=6.5cm] (label1) {\textbf{a) Successive equivalence queries and results}};

  \draw[very thick] (n1.south) -- (n2.north) node[midway, above left, text width=2.5cm] {\small{Counter-example $E_1$}};
  \draw[very thick] (n2.south) -- (n3.north) node[midway, above left, text width=2.5cm] {\small{Counter-example $E_2$}};

  \draw[gray] (n3.south) -- (n6.north) node[midway, above left, text width=2cm] {\small{Counter-example}};
  \draw[very thick] (n3.south) -- (n7.north) node[midway, above right, text width=1cm] {\small{Yes}};

  \draw[gray] (n1.south) -- (n4.north) node[midway, above right, text width=1cm] {\small{Yes}};
  \draw[gray] (n2.south) -- (n5.north) node[midway, above right, text width=1cm] {\small{Yes}};

  \node[draw, fill = green!10, rectangle, below right = -3mm and 4cm of n1.west, text width = 6.5cm, align=left, inner sep=5pt] 
  (d1) {\dOneText};
  \node[draw, fill = green!10, rectangle, below = of d1, text width = 6.5cm, align=left, inner sep=5pt, align = left, inner sep = 5pt] 
  (d2) {\dTwoText};
  
  \node[draw, rounded rectangle, right = of d1] (d3) {Correction \#2};
  \node[draw, rounded rectangle, right = of d2] (d4) {Correction \#3};

  \draw ([yshift=20pt]d1.north) -- (d1.north) node[midway, right, text width=5cm]
        {$\psi_1 \equiv pre \wedge \varphi_1 \wedge \varphi_2 \implies post$};
  \draw (d1.south) -- (d2.north) node[midway, right, text width=5cm]
        {$\psi_2 \equiv pre \wedge \varphi_1 \wedge \varphi_2' \implies post$};
  \draw (d2.south) -- ([yshift=-20pt]d2.south) node[midway, right, text width=5cm] {$\psi_3 \equiv pre \wedge \varphi_1 \wedge \varphi_2'' \implies post$};
  \draw[dashed] (d1.east) -- (d3.west) {};
  \draw[dashed] (d2.east) -- (d4.west) {};

  \node[below=8mm of d2.south] {\textbf{b) Refinement steps and corrections suggested}};
\end{tikzpicture}
}
\caption{Steps of the verified feedback generation algorithm
  for the DP update in the faulty submission in Figure~\ref{fig:faultymatrix}.}
\label{fig:algo-steps}
\end{figure*}
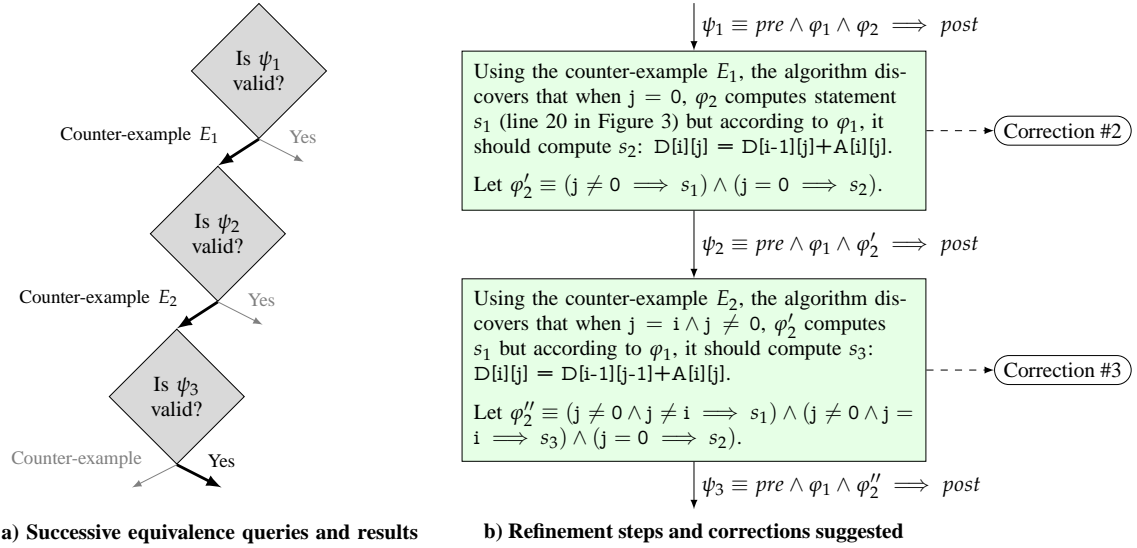

Extracting these features requires static analysis and syntactic pattern matching.
The most challenging part is identifying which array serves as a DP array.
An array which is defined in terms of itself is identified as a DP array
since in the DP recurrence, the DP array appears on both sides.
However, a student may use some temporary variables to store
intermediate results of DP computation and pass values across 
procedures. As explained in Section~\ref{sec:clustering}, we track data dependences
inter-procedurally. In Figure~\ref{fig:faultymatrix},
the array elements \code{D[i-1][j]} and \code{D[i-1][j-1]} are
passed to the procedure \code{max}. Through inter-procedural analysis,
our technique infers that the return value of \code{max} is indeed defined in
terms of these arguments and hence, \code{D} is defined in terms of
itself at line~\ref{subline:update}. Thereby, it
discovers that \code{D} is a DP array.

The submission in Figure~\ref{fig:faultymatrix} yields similar feature
values and is clustered along with the submission in Figure~\ref{fig:matrixpath}.
Note that our objective is not
to use clustering to distinguish between correct and incorrect submissions.
We therefore do not encode the exact nature of the initialization or update
of the DP array in the features.
Analyzing these is left to the verification and feedback phase.


\subsection{Verification and Feedback Phase}
\label{sec:feedback-example}

After clustering, suppose the instructor identifies the submission
in Figure~\ref{fig:matrixpath} as correct. Our objective is to verify
the submission in Figure~\ref{fig:faultymatrix} against it and suggest
fixes if any faults are found. For brevity, we will refer to the
submission in Figure~\ref{fig:matrixpath} as the \emph{reference} 
and the submission in Figure~\ref{fig:faultymatrix} as the \emph{candidate}.

By analyzing the sequence in which inputs are read, our technique infers that
the candidate uses two input
variables: an integer variable \code{n} and a 2D integer array \code{A},
where \code{n} is read first (line~\ref{subline:inp1})
and \code{A} second (line~\ref{subline:inp2}).
Their types respectively match the types of the
input variables \code{n} and \code{m} of the reference except
that the candidate uses a hardcoded array size for \code{A}. 
Both submissions use 2D integer DP arrays but the candidate hardcodes
array size of the DP array D also. While the reference can handle arbitrarily
large input matrices, the candidate can handle input
matrices only up to size $101 \times 101$. For smaller input matrices,
the reference is more space efficient than the candidate.
Our technique therefore emits correction \#1 in Figure~\ref{fig:feedback}
suggesting size declarations for \code{A} and \code{D}.

We check equivalence of matching code fragments of
the two submissions one-by-one. The matching code fragments are easy to
identify given the statement labels computed during feature extraction.
For our example, line~\ref{subline:update} of the candidate is an ``update'' statement
and lines~\ref{line:updt1}, \ref{line:updt2}, \ref{line:updt3} and \ref{line:updt4}
of the reference are also ``update'' statements.
Therefore, the top-level loop (say $L_2$)
at lines~\ref{subline:updtloopstart}--\ref{subline:update} of the candidate
matches the top-level loop (say $L_1$) at
lines~\ref{line:outerloopstart}--\ref{line:outerloopend} of the reference.
The question is whether they are equivalent. 

We check equivalence of the loop headers first. The input variables \code{n}
in both submissions correspond to each other and are not re-assigned before
they are used in the respective loop headers. Therefore, the loop headers
of $L_1$ and $L_2$ are equivalent. Thus, the corresponding loop indices 
are equal in each iteration.

To check equivalence of loop bodies, our algorithm formulates an \emph{equivalence
query} $\psi_1$ which asserts that in an (\code{i},\code{j})th iteration, if
the two DP arrays are equal at the beginning then they are equal at the end
of the iteration. The equivalence query is of the form:
\[
\psi_1 \equiv pre \wedge \varphi_1 \wedge \varphi_2 \implies post
\]
where 1)~$pre$ encodes the equality of DP arrays, loop indices and input variables
at the beginning of the iteration, and the lower and upper bounds on
the loop index variables in the reference,
2)~$post$ encodes the equality of DP arrays at the end of the iteration
(we syntactically check that input variables are not changed),
3)~$\varphi_1$ is a formula encoding the statements in the loop body of the reference,
and 4)~$\varphi_2$ is a formula encoding the statements in the loop body of the candidate.
Converting a loop-free sequence of statements into a formula is straightforward.
For example, an \code{if}-statement such as \code{if(p) x = e} is converted
to a guarded equality constraint $\code{p}: \code{x}' = \code{e}$
where $\code{x}'$ is a fresh variable.
The predicates in an \code{if-else} statement are propagated so as to make the
guards mutually disjoint and finally, the conjunction of all guarded equality
constraints is taken. We defer other technical details
to Section~\ref{sec:feedback-gen}.

As shown in Figure~\ref{fig:algo-steps}.a, the algorithm checks whether
$\psi_1$ is a logically valid formula. The SMT solver finds the
following counter-example $E_1$ which shows that the formula is not valid:
\begin{center}
\texttt{j} = \texttt{0}, \texttt{dp[i][j]} = \texttt{1},
\texttt{D[i][j]} = \texttt{2}, \texttt{m[i][j]} = \texttt{A[i][j]} = \texttt{1},\\
\texttt{dp[i-1][j]} = \texttt{D[i-1][j]} = \texttt{0},
\texttt{dp[i-1][j-1]} = \texttt{D[i-1][j-1]} = \texttt{1}
\end{center}
Following the usual convention, we use $=$ as equality in formulae and use \text{==} as
equality in code. Similarly, $\neq$ and \text{!=} denote disequality symbols in
formulae versus code.

Our algorithm, called \emph{counter-example guided feedback generation} algorithm,
uses $E_1$ to localize the fault in the candidate. It first identifies which
guards are satisfied by the counter-example in the candidate and the reference,
and whether they are equivalent.
The guard $\texttt{j}=\texttt{0}$ is satisfied in $\varphi_1$ and
the implicit guard $true$ for line~\ref{subline:update} is satisfied
in $\varphi_2$. Since they are not equivalent, the algorithm
infers that the faulty submission is missing a condition. On the contrary,
if the guards turn out to be equivalent, the fault is localized to
the assignment statement. It then derives
a formula $\varphi_2'$ given in Figure~\ref{fig:algo-steps}.b which
lets the candidate compute line~\ref{subline:update} under the guard
\texttt{j!=0} and makes it compute \texttt{D[i][j] = D[i-1][j]+A[i][j]}
under \texttt{j==0}. This assignment statement is obtained from the
assignment at line~\ref{line:updt1} under the guard \texttt{j==0}
of the reference in Figure~\ref{fig:matrixpath} by substituting the variables
from the candidate. The algorithm records this refinement in the form
of correction \#2 of Figure~\ref{fig:feedback}.

As shown in Figure~\ref{fig:algo-steps}.a, it checks validity of
$\psi_2 \equiv pre \wedge \varphi_1 \wedge \varphi_2' \implies post$ obtained
by replacing $\varphi_2$ (the encoding of candidate's loop body) by
$\varphi_2'$ (defined in Figure~\ref{fig:algo-steps}.b).
This results in a counter-example $E_2$ using which the algorithm
discovers the missing case of \texttt{j==i} and generates correction \#3
of Figure~\ref{fig:feedback}. For brevity, we do not show the counter-example $E_2$.
A refined equivalence query $\psi_3$
shown in Figure~\ref{fig:algo-steps}.b is computed.
As shown in Figure~\ref{fig:algo-steps}.a, this formula is valid
and establishes that the faults in the candidate can be fixed
using the synthesized feedback.

The input and initialization parts of the two submissions are found to be
equivalent. In our experiments, we observed certain
repeating iterative patterns. The computation of a maximum over an array in 
lines~\ref{subline:max1}--\ref{subline:max2} in Figure~\ref{fig:faultymatrix} is
one such example. We encode syntactic
patterns to lift these to certain predefined functions. We define \code{\_max} 
which takes the first and the last elements of a contiguous array segment
as arguments and returns the maximum over the array segment.
In Figure~\ref{fig:faultymatrix}, the output expression in terms of
\code{\_max} is \code{\_max(D[n-1][0], D[n-1][99])} and
in Figure~\ref{fig:matrixpath}, the
output is \code{\_max(dp[n-1][0], dp[n-1][n-1])}. A syntactic
comparison between the two leads to correction \#$4$ in Figure~\ref{fig:feedback}.

\section{Technical Details}
\label{sec:technicaldetails}

We now explain our approach for clustering submissions,
and the algorithm for verified feedback generation.

\subsection{Clustering by Solution Strategy}
\label{sec:clustering}

The first phase of our technique is to cluster submissions by
the solution strategy so that each cluster can be analyzed separately.

\subsubsection{Feature Design}
\label{sec:design}

Section~\ref{sec:clustering-example} has already introduced the features
of a submission that we use. 
Typically, in machine learning, a large number of features are obtained
and then the learning algorithm finds the important ones (called feature selection).
In our case, since the domain is well-understood, we design a small number of
suitable features that provide
enough information about the solution strategy.

In particular, we cluster two submissions together if
1) they use the same type and dimensions for the DP
arrays, 2) either both use DP arrays distinct from the input arrays or not,
and 3) there is a one-to-one correspondence between top-level loops
which contain DP update statements --- the loops should have the same
depth, direction and the DP array element being updated.
Two submissions in the same cluster can differ in all other aspects.

The rationale behind these features is simple: Checking equivalence of
two submissions which use the same types of DP arrays and similar DP update
loops is easier than if they do not share these properties. For example,
the subset-sum problem can be solved by using either a boolean DP array
or an integer DP array, but the two implementations are hard to
compare algorithmically. Recall the matrix path problem stated in the Introduction.
Consider a submission which traverses the matrix from top-to-bottom
and another which traverses it from bottom-to-top. Using one to validate
the other is difficult and perhaps, even undesirable.
The features of DP update loops will prevent
these submissions from being part of the same cluster.
Imposing further restrictions (by adding more features)
can make verification simpler but will increase the burden on the instructor
by creating additional clusters. 

The feature 4.c described in Section~\ref{sec:clustering-example} requires a
bit more explanation. We want to get the DP array element
being updated inside each loop containing a DP update statement.
If two submissions use different names for DP arrays and loop indices, we cannot
compare them. To compare them across submissions,
our technique uses canonical names for them: \code{dp} for the DP array
and loop indices \code{i}, \code{j}, \code{k}, etc. from the outer to inner loops.
If a submission uses multiple DP arrays then we assign subscripts to \code{dp}.

\subsubsection{Feature Extraction}
\label{sec:extraction}

Identifying input statements and variables is simple. We look for the common
C library functions like \code{scanf}. The case of output statements is
similar. A variable \code{x} is identified as a loop index variable if
1) \code{x} is a scalar variable,
2)~\code{x} is initialized before the loop is entered, 
3)~\code{x} updated inside the loop and
4)~\code{x} is used in the loop guard. Identifying DP arrays requires
more subtle analysis discussed below.
We call DP arrays, input variables and loop indices in a submission
as \emph{DP variables}. All other variables are called temporary variables.

To eliminate the use of a temporary variable \code{x} at a control
location $l$, we compute a set of guarded expressions
\[\{\code{g}_{\code{1}}:
\code{e}_{\code{1}} \text{,} \ldots \text{,}\; \code{g}_{\code{n}}:
\code{e}_{\code{n}}\}
\]
where the guards and expressions are defined only over DP variables,
and the guards are mutually disjoint.
We denote this set by
$\Sigma(l,\code{x})$ and call $\Sigma$ the \emph{substitution store}.
Semantically, if $\code{g}_{\code{k}}: \code{e}_{\code{k}} \in
\Sigma(l,\code{x})$ then \code{x} and $\code{e}_{\code{k}}$ evaluate to the same
value at $l$ whenever $\code{g}_{\code{k}}$ evaluates to true at $l$. 
The substitution store $\Sigma$ is lifted in a natural manner
to expressions and statements.  
For instance, for an assignment statement $s \equiv \code{x = e}$,
$\Sigma(l,s) = \{\code{g}_\code{1} : \code{x = e}_\code{1}\text{,}
\ldots\text{,}\; \code{g}_\code{n} : \code{x = e}_\code{n}\}$ where 
$\{\code{g}_{\code{1}}: \code{e}_{\code{1}} \text{,} \ldots \text{,}\;
\code{g}_{\code{n}}: \code{e}_{\code{n}}\} = \Sigma(l,\code{e})$.

Gulwani and Juvekar~\cite{bounds-analysis} developed an \emph{inter-procedural}
backward symbolic execution 
algorithm to compute symbolic bounds on values of expressions. While we are
not interested in the bounds, the equality mode of their algorithm
suffices to compute substitution stores. 
We refer the reader to~\cite{bounds-analysis} for the details.

To determine whether a statement $s$ at location $l$
is an initialization or an update statement,
we perform pattern matching over $\Sigma(l,s)$.
If the same array appears on both sides
of an assignment statement then the array is identified as a DP array
and the statement is labeled as an update statement. A statement where the LHS is a DP
array and RHS is an input variable or a constant is labeled as an initialization
statement. In $\Sigma(l,s)$, the 
temporary variables in $s$ are replaced by the guarded expressions from the
substitution store. This makes the labeling part of our tool robust even in
presence of temporaries and procedure calls.  For example, suppose we have
\code{t = x[i-1]; x[i] = t;}. The second statement can be identified as an
update statement through pattern matching only if we substitute \code{x[i-1]} in
place of \code{t} on the RHS of the statement.
In general, $\Sigma(l,s)$ may contain multiple
guarded statements.  If $\Sigma(l,s) =
\{s_1,\ldots,s_n\}$, we require that all of
$s_1,\ldots,s_n$ satisfy the same pattern and get the same label.
Extracting the feature values is now straightforward.

\subsubsection{Clustering and Identifying Correct Submissions}
\label{sec:presenting}

All our features are discrete valued. Therefore, our clustering algorithm
is very simple and works directly by checking equality of feature values.
Once the clustering is done, we ask the instructor to identify a correct
submission from each cluster. To reduce instructor's efforts,
we can employ some heuristics to rank candidates in a cluster
and present them one-by-one to the instructor.
For example, we can use
a small set of tests or majority voting on some other features 
of submissions like the loop bounds of update loops.

The instructor can accept a submission as correct or add a modified
version of an existing submission. If none of this is possible, the instructor
can write a correct solution similar to the solutions in the cluster.
If a new submission is added, we perform clustering again.
The instructor may have correct solutions from
a previous offering of the course if the assignment is repeated
from a previous offering. The instructor can
add them to the dataset even before we apply clustering to the
submissions.

\newlength{\oldtextfloatsep}\setlength{\oldtextfloatsep}{\textfloatsep}

\subsection{Verified Feedback Generation}
\label{sec:feedback-gen}

Once the submissions are clustered and the instructor has identified
a valid submission for each cluster, we proceed to the verified
feedback generation phase. We check semantic equivalence
of a submission from a cluster (called the \emph{candidate}) with
the instructor-validated submission from the same cluster
(called the \emph{reference}). 

\subsubsection{Variable and Control Correspondence}
\label{sec:bb}

Program equivalence checking is an undecidable problem.
In practice, a major difficulty is establishing correspondence between
variables and control locations of the two programs~\cite{narasamdya2005finding}.
We exploit the analysis information computed during feature extraction to solve
this problem efficiently.

Let $\sigma$ be a one-to-one function, called a \emph{variable map}.
$\sigma$ maps the input variables and DP arrays of the reference to
the corresponding ones of the candidate. To obtain a variable map,
the input variables of the two submissions are matched by considering
the order in which they are read and their types.
The DP arrays are matched based on their types. If there are multiple
DP arrays with the same type in both submissions then all type-compatible
pairs are considered. This generates a set of potential variable maps and
equivalence checking is performed for each variable map separately.
The one which succeeds and produces the minimum number of corrections is used
for communicating feedback to the student.
In equivalence checking,
we eliminate the occurrences of temporary variables using the substitution
store computed during feature extraction.
We therefore do not need to derive correspondence between temporary
variables -- which simplifies the problem greatly.

The feature extraction algorithm labels the
input, initialization, update and output statements of a submission.
We refer to these statements as \emph{labeled statements}.
The labeled statements give an easy way to establish control correspondence
between the submissions. We now use the notion of top-level statements defined
in Section~\ref{sec:clustering-example}.
Let $\hat{R} = [s_1^1,\ldots,s_1^k]$ be the list of all top-level statements of the
reference such that 1) each statement in $\hat{R}$ contains at least one labeled
statement and 2) the order of statements in $\hat{R}$ is consistent with their
order in the reference submission. It is easy to see that the top-level statements in
a submission are totally ordered. 
Let $\hat{C} = [s_2^1,\ldots,s_2^n]$ be the similar list for the candidate submission.
Without loss of generality, from now on, we assume that
there is only one DP array in a submission
and the top-level statements are (possibly nested) loops.

A (top-level) loop in $\hat{R}$ or $\hat{C}$ may contain multiple statements
which have different labels. For example, a loop may read the input and
also update the DP array. We call it a \emph{heterogeneous} loop.
If a loop reads two different input variables then also we call it
a heterogeneous loop. Heterogeneous loops make it
difficult to establish control correspondence between
the statement lists $\hat{R}$ and $\hat{C}$.
Fortunately, it is not difficult to canonicalize the statement lists using 
\emph{semantics-preserving} loop transformations, well-known in the compilers
literature~\cite{dragon-book}. Our algorithm first does loop splitting
to split a heterogeneous loop into different homogeneous loops.
It then does loop merging to coalesce different loops operating on the
same variable. Specifically, it merges two loops reading 
the same input array. It also merges loops performing
initialization to the same DP array. 
During merging, we ensure that there
is no loop in between the merged loops such that
it reads from or writes to the same variable or array as the merged loops.
In our experience, in most cases,
these transformations work because loops reading inputs or performing
initialization of DP arrays do \emph{not} have loop-carried dependences
or ad-hoc dependences between loops.

In contrast, by definition, loops performing DP updates do have
loop-carried dependences. We therefore do not attempt loop merging for
such loops. The feature $3$ in Section~\ref{sec:clustering-example}
tracks the number of loops containing DP updates.
Therefore, two submissions in the same cluster already have the
same number of loops containing DP updates. Thus, clustering helps
in reducing the variants that need to be considered during feedback generation.

Let $R$ and $C$ be the resulting statement lists for the reference and
candidate submissions respectively. If they have the same length and
at each index $i$, the $i$th loops in the two lists 1) operate on the
variables related by a variable map $\sigma$, 2) the statements operating
on the variables carry the same labels and 3) the loops have the
same nesting depth and directions then we get
the \emph{control correspondence} $\pi: R \rightarrow C$.
If our algorithm fails to compute variable or control correspondence for
the candidate then it exits without generating feedback, implicitly
delegating it to the instructor.

\subsubsection{Equivalence Queries}

Let $s_1'$ and $s_2'$ be the top-level loops from the reference
and the candidate such that $\pi(s_1') = s_2'$. We first use the
substitution map computed during feature extraction to eliminate
temporary variables and procedure calls in $s_1'$ and $s_2'$ by
equivalent guarded expressions over only DP arrays, loop indices
and input variables. Let $s_1 = \Sigma(l_1,s_1')$ and
$s_2 = \Sigma(l_2,s_2')$ where $l_1$ and $l_2$ are control locations
of $s_1'$ and $s_2'$.

We formulate an equivalence query $\Phi$ for the iteration spaces of $s_1$ and $s_2$.
Let $corr$ be the correspondence between the input variables, DP arrays, and
loop indices of $s_1$ and $s_2$ at the matching nesting depths.
We define $iter_1$ to be the range of the loop indices in $s_1$
and $guards_1$ to be the disjunction of all guards present in
the loop body of $s_1$. Similarly, we have $iter_2$ and $guards_2$
for $s_2$. The equivalence query $\Phi$ is defined as follows:
\[
\Phi \equiv corr \implies (iter_1 \wedge guards_1 \iff iter_2 \wedge guards_2)
\]

This query provides more flexibility than using direct syntactic
checking between the loop headers. For example, suppose $s_1$
is \code{for(i=1, i<=n, i++)\{$true$: s\}} and $s_2$ is
\code{for(i\ensuremath{'}=0, i\ensuremath{'}<=n, i\ensuremath{'}++)\{i\ensuremath{'} $> 0$: s\ensuremath{'}\}}. 
$s_1$ executes \code{s} for $1 \leq \code{i} \leq \code{n}$ and 
$s_2$ also executes $\code{s}'$ for $1 \leq \code{i}' \leq \code{n}$.
A syntactic check will end up concluding that $s_2$ executes one additional
iteration when \summary{i\ensuremath{'}} is $0$. But our equivalence
query establishes equivalence between the iteration spaces as desired.

The formulation of the query $\Psi$ to establish equivalence between loop
bodies of $s_1$ and $s_2$ is as discussed in Section~\ref{sec:feedback-example}.
Even though the submissions use arrays, we eliminate them from the queries.
A loop body makes use of only a finite number of symbolic array expressions.
We substitute each unique array expression in a query
by a scalar variable while encoding correspondence between the scalar variables
in accordance with the variable map $\sigma$.
We overcome some stylistic variations when the order of operands of
a commutative operation differs between the two submissions. For example,
say $s_1$ uses \code{x[i+j]} and $s_2$ uses \code{y[b+a]} such that
$\sigma(\code{x})= \code{y}$, $\sigma(\code{i}) = \code{a}$ and
$\sigma(\code{j}) = \code{b}$. The expressions \code{i+j}
and \code{b+a} are not identical under renaming but are equivalent due
to commutativity. To take care of this, we force a fixed ordering among variables
in the two submissions for commutative operators.
Sometimes, the instructor may include some constraints over input variables
as part of the problem statement. 
In the equivalence queries, our algorithm takes input constraints
into account and also adds array bounds checks. We omit these details
due to space limit.


\subsubsection{Counter-Example Guided Feedback Generation}

\begin{algorithm}[t]
\setlength{\oldtextfloatsep}{\textfloatsep}
\setlength{\textfloatsep}{0pt}
\small {
\SetAlgoNoLine
\DontPrintSemicolon
\SetInd{0.5em}{0.5em}
\KwIn{A list $Q = [(\Phi_1,\Psi_1),\ldots,(\Phi_k,\Psi_k)]$ of equivalence
  queries}
\KwOut{A list of corrections to the candidate submission}
\ForEach{$(\Phi_i,\Psi_i) \in Q$}{
  \label{line:outerloop}
  \If{$\exists \alpha \not\models \Phi_i$}{
  \label{line:startcheckphi}
            Suggest corrections to make the iteration spaces
         of the $i$th statements of the two submissions equal\;

    }
    \label{line:endcheckphi}
    Let $\Psi_i \equiv pre \wedge \varphi_1 \wedge \varphi_2 \implies post$\;
    $k \leftarrow 0$\;
    \Repeat{$k < \delta$}
           {
    \label{line:startrefineloop}
             $k \leftarrow k+1$\;
             \lIf{$\models \Psi_i$}{
	     \label{line:checkpsi}
               \textbf{break}
             }
             \Else    {
               Let $\alpha \not\models \Psi_i$ be a counter-example\;
               Let $g_1: s_1 \in \varphi_1$ and $g_2:s_2 \in \varphi_2$ s.t.           
               $\alpha \models g_1$ and $\alpha \models g_2$\;
	       \label{line:localize}
               \eIf{$\models pre \implies (g_1 \iff g_2)$}{
	       \label{line:guardeq}
                 $\varphi_2' \leftarrow \varphi_2[g_2:s_2/g_2:\hat{\sigma}(s_1)]$\;
		 \label{line:replacestmtcase1}
                 $\Psi_i \leftarrow \Psi_i[\varphi_2/\varphi_2']$\;
		 \label{line:replacequerycase1}
                 Suggest computation of $\hat{\sigma}(s_1)$ instead of $s_2$ under $g_2$\;
		 \label{line:suggest1}
               }
                   {
		   \label{line:startcase2}
                     $h_2 \leftarrow g_2 \wedge \hat{\sigma}(g_1)$; \phantom{a}
                     $h_2' \leftarrow g_2 \wedge \hat{\sigma}(\neg g_1)$\;
		     \label{line:h2def}
                     $\varphi_2' \leftarrow \varphi_2[g_2:s_2/h_2:\hat{\sigma}(s_1) \wedge
                       h_2':s_2]$\;
		     \label{line:replacestmtcase2}
                     $\Psi_i \leftarrow \Psi_i[\varphi_2/\varphi_2']$\;
		     \label{line:replacequerycase2}
                     Suggest computation of $\hat{\sigma}(s_1)$ instead of $s_2$ under $h_2$\;
		     \label{line:suggest2}
                   }
		   \label{line:endcase2}
             }
           }
	   \label{line:endrefineloop}
           \lIf{$k = \delta$}{
	   \label{line:thresh}
             Suggest a correction to replace $\varphi_2$ by $\hat{\sigma}(\varphi_1)$
	     \label{line:suggest3}
             }
  }
}
\caption{Algorithm {\sc GenFeedback}}
\label{algo:feedback}
\afterpage{\global\setlength{\textfloatsep}{\oldtextfloatsep}}
\end{algorithm}

Algorithm~\ref{algo:feedback} is our counter-example guided feedback generation
algorithm. Its input is a list $Q$ of equivalence queries where
each query $(\Phi_i,\Psi_i)$ corresponds to the $i$th statements in the
two submissions. $\Phi_i$ encodes the equivalence of iteration spaces
and $\Psi_i$ of the loop bodies. If the $i$th statements are not loops,
$\Phi_i$ is $true$ and $\Psi_i$ just checks equivalence of the loop-free
statements. The output of the algorithm is a list of corrections to
the candidate submission.

Algorithm~\ref{algo:feedback} iterates over the query list (line~\ref{line:outerloop}).
For a query $(\Phi_i,\Psi_i)$, it first checks whether $\Phi_i$ is (logically)
valid or not. If it is not then the algorithm suggests a correction to
make the iteration spaces of the $i$th statements (loops) of the
two submissions equal (lines~\ref{line:startcheckphi}-\ref{line:endcheckphi}). It then enters a refinement loop for $\Psi_i$ at
lines~\ref{line:startrefineloop}-\ref{line:endrefineloop}.

During each iteration of the refinement loop, it checks whether $\Psi_i$
is valid. If yes, it exits the loop (line~\ref{line:checkpsi}). Otherwise, it gets a
counter-example $\alpha$ from the SMT solver and finds the guarded statements
that are satisfied by $\alpha$. Let $g_1:s_1 \in \varphi_1$ and
$g_2:s_2 \in \varphi_2$ be those statements (line~\ref{line:localize}). The formulae
$\varphi_1$ and $\varphi_2$ correspond to the encodings of the loop bodies
of the reference and the candidate respectively.
Note that the conversion
of statements to guarded equality constraints (Section~\ref{sec:feedback-example}) ensures
that the guards within $\varphi_1$ and within $\varphi_2$ are pairwise disjoint.

Let $\hat{\sigma}$ be the variable map which is same as the
variable correspondence $\sigma$ but augmented with the correspondence
between loop indices at the same nesting depths for the $i$th statements.
The function $\hat{\sigma}$ is lifted in a straightforward manner to
expressions and assignments. 
The algorithm checks whether the guards $g_1$ and $g_2$ are equivalent
(line~\ref{line:guardeq}).
If they are then the fault must be in the assignment statement $s_2$.
It therefore defines $\varphi_2'$ by substituting $s_2$ by $\hat{\sigma}(s_1)$
in $\varphi_2$ (line~\ref{line:replacestmtcase1}) and refines $\Psi_i$ by replacing $\varphi_2$
by $\varphi_2'$ (line~\ref{line:replacequerycase1}). It suggests an appropriate correction for
the candidate submission (line~\ref{line:suggest1}).
The other case when the guards are not equivalent leads to the other
branch (lines~\ref{line:startcase2}-\ref{line:endcase2}). The algorithm now splits the guarded assignment
$g_2: s_2$ to make it conform to the reference under
$h_2 \equiv g_2 \wedge \hat{\sigma}(g_1)$, whereas, for
$h_2' \equiv g_2 \wedge \hat{\sigma}(\neg g_1)$, the candidate can continue
to perform $s_2$ (line~\ref{line:h2def}). It computes $\varphi_2'$ by
replacing $g_2:s_2$ by $h_2: \hat{\sigma}(s_1)$ and $h_2' : s_2$
(line~\ref{line:replacestmtcase2}). It then refines $\Psi_i$ by replacing $\varphi_2$ by $\varphi_2'$
(line~\ref{line:replacequerycase2}) and suggests an appropriate correction for the candidate submission
(line~\ref{line:suggest2}).
The refinement loop terminates when no more
counter-examples can be found (line~\ref{line:checkpsi}) and thus, progressively
finds \emph{all} semantic differences between $i$th statements of the two submissions. 

Each iteration of the refinement loop eliminates a semantic difference between
a pair of statements from the two submissions 
and the loop terminates after a finite number of iterations. 
In practice, giving a long list of corrections might not be
useful to the student if there are too many mistakes in the submission. A better
alternative might be to stop generating corrections after a threshold is reached.
We use a constant $\delta$ to control how many refinements should be attempted
(line~\ref{line:endrefineloop}). If this threshold is reached then the algorithm suggests a
total substitution of $\hat{\sigma}(\varphi_1)$ in place of $\varphi_2$
(line~\ref{line:thresh}). In our experiments, we used $\delta = 10$.

Due to the explicit verification of equivalence queries,
our algorithm only generates correct feedback. The feedback for the
declarations of the candidate are obtained by checking dimensions of
the corresponding variables according to $\sigma$.

\section{Implementation}
\label{sec:implementation}

We consider C programs for experimental evaluation.
We have implemented the source code analysis using the Clang front-end of
the LLVM framework~\cite{LLVM:CGO04} and use Z3~\cite{z3} for SMT solving.
We presently do not support pointer arithmetic.

In the pre-processing step, our tool performs
some syntactic transformations. It rewrites compound assignments
into regular assignments. For example, \code{x += y}
is rewritten to \code{x = x + y}. 
A code snippet of the form: \code{scanf("\%d",
\&a[0]); for (i = 1; i < n; i++) scanf("\%d", \&a[i]);}, where the input array
is read in multiple statements is transformed to use a single read statement. The
above snippet will be rewritten to \code{for (i = 0; i < n; i++) scanf("\%d",
\&a[i]);}. Sometimes, students read a scalar variable and then
assign it to an array element. Our tool eliminates the use of the scalar variable and
rewrites the submission so that the input is read directly into the array element.
Another common pattern is to read a sequence of input values into
a scalar one-by-one and then use it in the DP computation. For
example, consider the code snippet: \code{for (i = 0; i < n; i++) for (j = 0; j
  < n; j++) \{ scanf("\%d", \&x); dp[i][j] = dp[i-1][j] + x; \}}. It does
not use an array to store the sequence of input values. We
declare an array and rewrite the snippet to use it.
When feedback is generated for the submission,
an explanatory note about the input array is added.
In each of the syntactic
transformations, we ensure that the program semantics is not altered. 


Many students, especially beginners, write programs with convoluted conditional
control flow, and unnecessarily complex expressions. 
In addition, the refinement steps of our
counter-example guided feedback generation algorithm may generate
complex guards. To present {clear and concise feedback} even in the face
of these possibilities, in the post-processing step,
our tool simplifies guards in the feedback using the SMT solver. We use
Z3's tactics to remove redundant clauses, evaluate sub-expressions to
Boolean constants and simplify systems of inequalities.

\section{Experimental Evaluation}

\begin{table}[t]
  \centering
  \caption{Summary of submissions and clustering results.}
  {
  \begin{tabular}{lrrr}
    \toprule
    Problem & Total & Clusters with  & Clusters with \\
    & subs.	   & correct sub.	 & manually added\\
    & & & correct sub.\\
    
    \midrule
    \code{SUMTRIAN} & $1983$ & $78$ & $2$\\
    \code{MGCRNK}   & $144$ & $23$ & $3$\\
    \code{MARCHA1}  & $58$ & $4$ & $2$\\
    \code{PPTEST}   & $41$ & $2$ & $0$\\
    \midrule
    Total           & $2226$ & $107$ & $7$\\
    \bottomrule
  \end{tabular}
  }
  \label{tab:stats}
\end{table}

To assess the effectiveness of our technique, we collected submissions
to the following $4$ DP problems\footnote{\small{\url{http://www.codechef.com/problems/<problem-name>}}} on CodeChef:
\begin{enumerate}\denseitems
\item \code{SUMTRIAN} -- Described in the Introduction section.
\item {\code{MGCRNK}} --
Find a path from (1,1) to (N,N) in an N $\times$ N
matrix, so that the average of all integers in cells on the
path, excluding the end-points, is maximized. From each
cell, the path can extend to cells to the right or below.
\item {\code{MARCHA1}} -- The subset sum problem.
\item {\code{PPTEST}} -- The knapsack problem.
  \end{enumerate}

We selected submissions to these problems that implemented
an iterative DP strategy in the C language.
A user can submit solutions any number of times.
We picked the latest submissions from individual users. 
These represent their best efforts and can benefit from feedback.
We do not consider submissions that either do not compile or
crash on CodeChef's tests.
To enable automated testing on CodeChef, the
submissions had an outermost loop to iterate over test cases -- we
identified and removed this loop automatically before further analysis. 

Table~\ref{tab:stats} shows the number of submissions for each problem.
\code{SUMTRIAN} had the maximum number of submissions ($1983$) and
\code{PPTEST} had the minimum ($41$). There were a total of
$2226$ submissions from 
$1860$ students representing over $250$ institutions.
These submissions employ a wide range of coding idioms and many possible solution
approaches, both correct and incorrect. 
This is a fairly large, diverse and challenging set of submissions.

\subsection{Effectiveness of Clustering}

Our features were quite effective in clustering submissions by their solution
strategies. Since we do not include features representing low-level syntactic
or structural aspects of submissions, the clustering resulted in only a
few clusters for each problem, without compromising our ability to generate
verified feedback. Table~\ref{tab:stats} gives the number of
clusters. The number of clusters increased gracefully from the smallest
problem (by the number of submissions) to the largest one.
The smallest problem \code{PPTEST}
yielded only $2$ clusters for $41$ submissions, whereas,
the largest problem \code{SUMTRIAN} yielded $80$ clusters for $1983$ submissions.
Our manual evaluation revealed that in each cluster, the solutions were
actually following the same DP strategy. 

The small number of clusters reduces the burden on the instructor significantly.
Instead of evaluating $2226$ submissions separately, the instructor is required to
look at representatives from only $114$ clusters.
CodeChef uses test suites to classify problems into correct and incorrect. As a
simple heuristic, we randomly picked one of the submissions marked as
correct by CodeChef in each
cluster and manually validated it. As shown in Table~\ref{tab:stats},
this gave us correct representatives
for $107/114$ clusters across the problems. 
The remaining $7$ clusters
seemed to follow some esoteric strategies and we manually added 
a correct solution to each of them.

Clustering also helps the instructor get {a bird's eye view of
the multitude of solution strategies}.
For example, it can be used to find the most or least
popular strategy used in student submissions.
In \code{SUMTRIAN}, the most
popular strategy (with $677$ submissions)
was the one that traverses the matrix
rows bottom up, traverses the columns left to right and updates the element
$(\texttt{i},\texttt{j})$. 

\subsection{Effectiveness of Feedback Generation} 

\begin{table}[t]
  \centering
  \caption{Results of feedback generation.}
  {\small
  \begin{tabular}{@{}lrrrr@{}}
    \toprule
    Problem & Verified as		  & Corrections & Average & Unlabeled \\ 
    & correct ({\color{green}\checkmark}) &  suggested ({\color{red}\ding{55}}) & corrections & (\textbf{?})\\
    \midrule
    \code{SUMTRIAN} & $1049$ & $659$ & $3.3$ & $275$\\
    \code{MGCRNK}   & $61$   & $66$ & $6.8$ & $17$\\
    \code{MARCHA1}  & $9$    & $35$ & $10.3$ & $14$\\
    \code{PPTEST}   & $3$    & $29$ & $12.7$ & $9$\\
    \midrule
    Total           & $1122$ & $789$ & $4.3$ & $315$\\
    \bottomrule
  \end{tabular}
  }
  \label{tab:results}
\end{table}

Our tool verifies a submission from a cluster against the manually validated
or added correct submission from the same cluster. Table~\ref{tab:results} shows
the number of
1) submissions verified as correct ({\color{green}\checkmark}),
2) submissions for which faults were identified and corrections suggested ({\color{red}\ding{55}}) and
3) submissions which our algorithm could not handle (?).
Across the problems, $1122$ submissions amounting to
$50$\% were verified to be correct, with
the maximum at $53$\% for \code{SUMTRIAN} and the minimum at $7$\% for \code{PPTEST}.
For a total of $789$ submissions amounting to $35$\%, some corrections
were suggested by our tool. The maximum percentage of submissions with
corrections were for \code{PPTEST} at $71$\% and the minimum was $33$\%
for \code{SUMTRIAN}. Many submissions had
multiple faults. Table~\ref{tab:results} shows the average number
of corrections over faulty submissions for each problem.
\code{PPTEST} required
the maximum number of corrections of $12.7$ on average.
In all, our tool succeeded in either verifying or
generating verified feedback for $85$\% submissions.

For the remaining $315$ ($15$\%) submissions, our tool could neither
generate feedback nor verify correctness. These submissions need manual evaluation.
\code{MARCHA1} had the maximum
percentage of unlabeled submissions at $24$\% and \code{MGCRNK} had the
minimum at $12$\%. These arise either because the SMT solver times out
(we kept the timeout of $3$s for each equivalence query), or
due to the limitations of the verification algorithm or the implementation.

These results on the challenging set of DP submissions are encouraging and
demonstrate effectiveness of our methodology and technique.
Even if we assume that all $315$ unhandled submissions are
faulty, we could generate verified feedback for $71$\% faulty submissions.
In comparison, on a set of \emph{introductory} programming assignments,
Singh et al.~\cite{autograder}
report that $64$\% of faulty submissions could be fixed using
\emph{manually provided error models}.
Our counter-example guided feedback generation
technique guarantees
correctness of the feedback. In addition,
we would have liked to communicate the feedback to the students and
assess their responses. Unfortunately, their contact details were not
available to us. 

\paragraph{Diversity of Feedback and Personalization}

\begin{figure}
  \centering
\setlength{\belowcaptionskip}{-15pt}
\begin{tikzpicture}[scale=0.7]
    \pie{0.7/I only, 2.9/O only, 7.9/Correct, 7.9/I\&U, 7.9/I\&U\&O, 16.4/U only, 26.4/U\&O, 30/Others}
\end{tikzpicture}
\caption{Distribution of submissions in a cluster of \code{SUMTRIAN}
  by the type of feedback.
}
\label{fig:pie}
\end{figure}
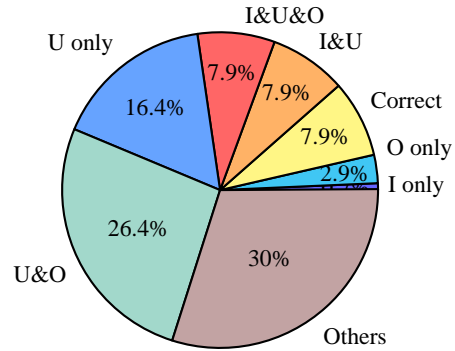

The feedback propagation approaches~\cite{HuangAIED,DBLP:conf/icml/PiechHNPSG15}
suggest that the same feedback text written by the instructor can be
propagated to all submissions within a cluster. We found that this is not
practical and the submissions within the same cluster require heterogeneous
feedback. Figure~\ref{fig:pie} shows the distribution of submissions
in a cluster of \code{SUMTRIAN} by
the type of feedback. 
We only highlight feedback over the logical components of a submission:
initialization (I), update (U) and output (O). Feedback related to type declarations
and input statements (possibly, in conjunction with feedback on the 
logical components) is summarized under the category ``Others''.

While only $7.9$\%
submissions were verified to be correct, $20$\% submissions had faults
in one of the logical components of the DP strategy:
initialization ($0.7$\%), update ($16.4$\%) and output ($2.9$\%).
As shown in Figure~\ref{fig:pie}, a large percentage of submissions had
faults in two logical components, and $7.9$\% had them in all three components.
$30$\% of the submissions were in the others category.
Clearly, it would be difficult for the instructor to predict faults in
other submissions in a cluster by looking only at some submissions in the cluster
and write feedback applicable to all.
We do admit that Figure~\ref{fig:pie} is based on our clustering approach
and other approaches may yield different clusters. Even then, the clusters
would be correct only in a \emph{probabilistic} sense and the verification phase,
we suggest, would add \emph{certainty} about correctness of feedback.

\begin{table}[t]
  \centering
  \caption{Submissions by faulty components.}
  \small{
  \begin{tabular}{@{}lrrrr@{}}
    \toprule
    Faulty
    comp. & {\code{SUMTRIAN}} & {\code{MGCRNK}} & {\code{MARCHA1}} & {\code{PPTEST}}\\ 
    \midrule
    I only & $36$ & $15$ & $0$ & $0$\\
    U only & $229$ & $7$ & $5$ & $2$\\
    O only & $31$ & $1$ & $6$ & $0$ \\
    I\&U & $29$ & $18$ & $2$ & $8$\\
    I\&O & $10$ & $0$ & $1$ & $0$ \\
    U\&O & $97$ & $1$ & $2$ & $0$ \\
    I\&U\&O & $30$ & $0$ & $11$ & $0$ \\
    Others & $197$ & $24$ & $8$ & $19$\\
    \midrule
    Total & $659$ & $66$ & $35$  & $29$\\
    \bottomrule
  \end{tabular}
  }
  \label{tab:component-type}
\end{table}

Our technique generated personalized feedback depending on which 
components of a submission were faulty. Table~\ref{tab:component-type} shows the
number of submissions by the faulty components. Across the problems,
\code{PPTEST} had the maximum percentage $53.7$\%
of submissions requiring corrections to multiple logical components and
\code{SUMTRIAN} had the minimum percentage $17.5$\%. The most common
faulty components varied across problems.

\paragraph{Types of Faults Found and Corrected}

Our tool found a wide range of faults and suggested appropriate
corrections for them. This is made possible by availability of a correct
submission to verify against and the ability of our verification algorithm
to refine the equivalence queries to find all faults.
The faults found and corrected include: incorrect loop headers, initialization mistakes including
missing or spurious initialization, missing cases in the DP recurrence, errors
in expressions and guards, incorrect dimensions, etc.  

\pgfplotstableread[col sep = comma]{feedbackSize.csv}\loadedtable

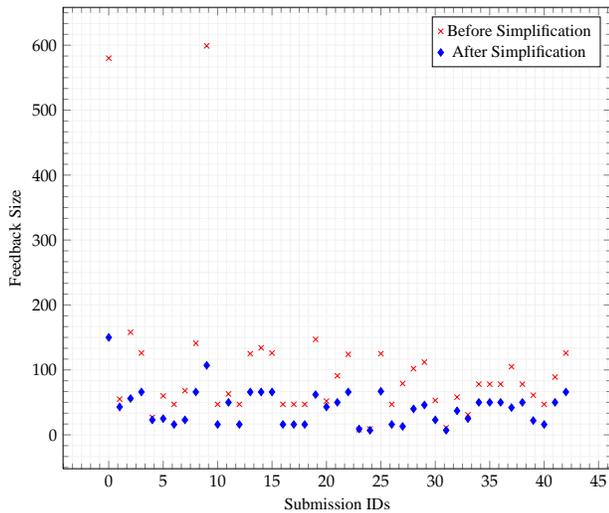
\begin{figure}[t]
\centering
\setlength{\belowcaptionskip}{-15pt}
\begin{tikzpicture}[scale=0.7]
    \begin{axis}[
        xlabel=Submission IDs,
        ylabel=Feedback Size,
        grid=both,
        grid style={line width=.1pt, draw=gray!10},
        major grid style={line width=.1pt,draw=gray!10},
        minor tick num=5,
      ]
        \addplot[only marks, mark=x, color=red] table [x=Submission, y=NoOpt, col sep=comma] {\loadedtable};
        \addlegendentry{Before Simplification};
        \addplot[only marks, mark=diamond*, color=blue] table [x=Submission, y=Opt, col sep=comma] {\loadedtable}; 
        \addlegendentry{After Simplification};
    \end{axis}
\end{tikzpicture}
\caption{Effect of simplification on feedback size for \code{MGCRNK}}
\label{fig:simplification}
\end{figure}

\paragraph{Conciseness of Feedback}

To reduce the size of formulae in the generated feedback, we perform
simplifications outlined in Section~\ref{sec:implementation}.
We measure the effectiveness of the
simplifications by disabling them and using the sum of AST sizes (\#nodes 
in the AST) of the guards in our feedback text as 
\emph{feedback size}.
Figure~\ref{fig:simplification} shows the impact of the
simplifications on feedback size in the
case of \code{MGCRNK} by plotting submission IDs versus feedback size.
The figure excludes cases where simplification
had no impact on feedback size. 
Simplifications ensured that the feedback size was at most $150$, and
$42.1$
on average. Without simplifications, the maximum feedback size was
$599$. Simplifications, where applicable, reduced feedback size by
$63.1\%$  on an average across the problems.

\subsection{Comparison with CodeChef}
\label{sec:codechef}

Our tool was able to verify $12$ submissions as correct that were tagged by
CodeChef as incorrect. This was surprising because
CodeChef uses tests which should not produce such \emph{false positives}.
On investigation, we found that the program logic was indeed correct, as verified by
our tool. The faults were localized to output formatting, or in
custom input/output functions. Understandably, black-box testing used by CodeChef
cannot distinguish between formatting and logical errors.
However, being able to distinguish between these
types of faults would save time for the students. Our tool finds logical faults
but not formatting errors.

Due to the incompleteness of testing, CodeChef did not identify all faulty
submissions (\emph{false negatives}). This can hurt students since they may not realize
their mistakes. We checked the cases when CodeChef tagged a submission as
correct but our tool issued some corrections.  For $64$ submissions, our tool
identified that the submissions were making spurious initializations to the DP
array.  For $112$ submissions, our tool identified that the DP udpate was
performed for additional iterations than required and generated feedback to fix
the bounds of loops containing update statements.  Importantly, our tool
detected \emph{out-of-bounds array accesses} in $99$ submissions, and suggested
appropriate corrections. In $265$ distinct submissions, our tool was
able to identify one or more of the faults described above, whereas CodeChef
tagged them as correct! Thus, our static technique has a
qualitative advantage over the test-based approach of online judges.

\subsection{Performance}

We ran our experiments on
an Intel Xeon E5-1620 3.60 GHz machine with 8 cores and 24GB
RAM. Out tool runs only on a single core. On an average, our tool generated
feedback in $1.6$s including the time for clustering and
excluding the time for identifying correct submissions manually.

\subsection{Limitations and Threats to Validity}

Our technique fails for submissions that have
loop-carried dependencies over scalar variables apart from the loop index variables,
submissions that use auxiliary arrays and submissions for which pattern matching fails
to label statements. We inherit the limitations of SMT
solvers in reasoning about non-linear constraints and program expressions with
undefined semantics, such as division by 0. Most of the unhandled cases
arise from these limitations.

Our approach cannot
suggest feedback for errors in custom input/output functions, output formatting,
typecasting, etc. Our approach may provide spurious feedback 
enforcing stylistic conformance with the instructor-validated submission.
For example, if a submission starts indexing into arrays from position $1$
but the instructor-validated submission indexes from position $0$, our tool
generates feedback requiring the submission to follow $0$ based indexing.
This may correct some misconception about array indexing that the student may
have. Nevertheless, these differences can be either compiled away during
pre-processing or through SMT solving with additional annotations.
We will investigate these in future. Finally,
our implementation currently handles only a frequently used subset of C
constructs and library functions.

There can be faults in our implementation that might have affected our
results. To address this threat, we manually checked the feature values and
feedback obtained, and did not encounter any error. 
Threats to external validity arise because
our results may not generalize to other problems and submissions.
We mitigated this threat by drawing upon submissions from
more than $1860$ students on $4$ different problems.
While our technique is able to handle most constructs that
introductory DP coursework employs, further studies are required to
validate our findings in the case of other problems. In Section~\ref{sec:codechef},
we compared our tool
with the classification available on CodeChef. The tests used
by CodeChef are not public and hence, we cannot ascertain their quality.
By improving the test suites, some false negatives of CodeChef
may disappear but black-box testing will not be able to distinguish between
logical faults and formatting errors (discussed in Section~\ref{sec:codechef}).

\section{Related Work}

\paragraph{Program Representations and Clustering}

In order to cluster submissions effectively, we need strategies
to represent both the syntax and semantics of programs.
Many clustering approaches use only edit distance between
submissions~\cite{gross2012cluster,Rivers},
while others use edit distance along with test-based
similarity~\cite{HuangAIED,codewebs,overcode}.
We use neither of these.
Glassman et al.~\cite{Glassman:2014:FEC:2556325.2567865} advocate a 
hierarchical technique where the submissions are first clustered using
high-level (abstract) features and then using low-level (concrete) features.
An interesting recent direction is to use deep learning to compute and use
vector representations of programs~\cite{peng2015building,DBLP:conf/icml/PiechHNPSG15,mou2016convolutional}. Peng et al.~\cite{peng2015building}
propose a pre-training technique to automatically
compute vector representations of different AST nodes which is then
fed to a tree-based convolution neural network~\cite{mou2016convolutional} for a classification
task. Piece et al.~\cite{DBLP:conf/icml/PiechHNPSG15} propose a recursive neural network to
capture both the structure and functionality of programs. The functionality
is learned using input-output examples. But the class of programs considered
in~\cite{DBLP:conf/icml/PiechHNPSG15} is very simple. It only handles
programs which do not have any variables. 

Since our experiments were focused on iterative DP solutions,
we designed features that capture the DP strategy. 
The above approaches are more general but
unlike us, they may not put the submissions in Figure~\ref{fig:matrixpath}
and~\ref{fig:faultymatrix} in the same cluster.
Our algorithm
extracts features in the presence of temporary variables and procedures, and
might be useful in other contexts as well. 

\paragraph{Feedback Generation and Propagation}

The idea of comparing instructor provided
solutions with student submissions appears in~\cite{laura}.
It uses graph representation and transformations for comparison of Fortran programs.
Xu and
Chee~\cite{transformXu} use richer graph representations for
object-oriented programs. 
Rivers and Koedinger~\cite{Rivers} use
edit distance as a metric to compare graphs and generate feedback.
Gross et al.~\cite{gross2012cluster} cluster student solutions by structural
similarity and perform syntactic comparisons with a known correct solution
to provide feedback. Feedback generated by pattern matching may not always be correct.
In contrast, we generate \emph{verified}
feedback but for the restricted domain of DP.

Alur et al.~\cite{alurautomata} develop a technique to
automatically grade automata constructions using a pre-defined set of corrections.
Singh et al.~\cite{autograder} apply sketching based synthesis to provide
feedback for introductory programming assignments.
In addition to a reference implementation, the
tool takes as input an error model in the form of correction rules.
Their error model  is 
too restrictive to be adapted to our setting 
that requires more sophisticated repairs and that too for a more challenging
class of programs. 
Gulwani et al.~\cite{perfFeed} address the orthogonal issue of
providing feedback to address performance issues, while 
Srikant and Aggarwal~\cite{SrikantKDD} use machine learning to
assess coding quality of prospective employees and do not provide
feedback on incorrect solutions.

The idea of exploiting the common patterns in DP programs has been
used by Pu et al.~\cite{PuOopsla} but for synthesis of DP programs.
The clustering-based approaches~\cite{HuangAIED,DBLP:conf/icml/PiechHNPSG15}
propagate the instructor-provided feedback 
to all submissions in the same cluster, whereas we
generate personalized and verified feedback for each submission in a cluster
separately. 
OverCode~\cite{overcode} also performs clustering of submissions
and provides a visualization technique to assist the instructor in
manually evaluating the submissions. 

\paragraph{Program Repair and Equivalence Checking}

Genetic programming has been used to automatically
generate program repairs~\cite{arcuri2008,debroy2010using,legoues-tse2012}.
These approaches
are not directly applicable in our setting as the search space of mutants is
very large. Further, GenProg~\cite{legoues-tse2012} relies on redundancy present in
other parts of the code for fixing faults. This condition is not met in our
setting.
Software transplantation~\cite{DBLP:conf/wcre/HarmanLW13,DBLP:conf/issta/BarrHJMP15}
transfers functionality from
one program to another through genetic programming and slicing.
Prophet~\cite{DBLP:conf/popl/LongR16} learns a probabilistic,
application independent model of correct code from existing patches,
and uses it to rank repair candidates from a search space.
These are generate-and-validate approaches which
rely on a test suite to validate the changes. In comparison,
we derive corrections for a faulty submission by program equivalence checking
with a correct submission.

Konighopher et. al.~\cite{konighofer2011automated} present a repair
technique using
reference implementations. 
Their fault model is restrictive and only considers faulty RHS.
Many approaches rely on program specifications for
repair, including contracts~\cite{pei2011code,wei10}, 
LTL~\cite{jobstmann2005program}, assertions~\cite{costaware} and pre-post
conditions~\cite{gopi2011,ball2012,he2004}.  
Recent approaches that use tests to infer specifications and propose repairs
include SemFix~\cite{nguyen2013}, 
MintHint~\cite{minthint}, DirectFix~\cite{directfix} and Angelix~\cite{angelix}.
These approaches use synthesis~\cite{jha2010oracle}, symbolic
execution~\cite{klee} and partial MaxSAT~\cite{z3} respectively. Both DirectFix
and Angelix use partial MaxSAT but Angelix extracts more lightweight
repair constraints to achieve scalability.
SPR~\cite{staged-program-repair} uses parameterized transformation schemas
to search over the space of program repairs.
In contrast,
we use instructor-validated submissions and a combination of
pattern matching, static analysis and SMT solving.
%


Automated equivalence checking between a program and its optimized version
has been studied in translation validation~\cite{pnueliEquiv, neculaEquiv,
  barretEquiv}.
Partush and Yahav~\cite{yahavEquiv} design an abstract interpretation
based technique to check equivalence of a program and its patched
version. 
In comparison, our technique performs equivalence
check between programs written by {different individuals} independently.

All these approaches are designed for developers and deal
with only one program at a time. Our tool
targets iterative DP solutions written by students
and works on a large number of submissions simultaneously. It combines
clustering and verification to handle both the scale and variations in
student submissions.

\section{Conclusions and Future Work}

We presented semi-supervised verified feedback generation 
to deal with both scale and variations in student submissions,
while minimizing the instructor's efforts and ensuring feedback quality.
We also designed a novel counter-example guided feedback generation algorithm.
We successfully demonstrated the effectiveness of our technique on
$2226$ submissions to $4$ DP problems. 

Our results are encouraging and suggest that the combination of
clustering and verification can pave way for practical feedback generation
tools. There are many possible directions to improve clustering and
verification by designing sophisticated algorithms.
We plan to investigate these for more problem domains.

\bibliographystyle{abbrv}
\bibliography{paper}

\end{document}